\documentclass[aps,prl,twocolumn,amsmath,showpacs,superscriptaddress]{revtex4-1}
\usepackage{bm}
\usepackage{graphicx}
\begin{document}
\setlength{\arraycolsep}{2pt}
\title{Efficient entanglement criteria beyond Gaussian limits using Gaussian measurements}
\author{Hyunchul Nha}
\email{hyunchul.nha@qatar.tamu.edu }
\affiliation{Department of Physics, Texas A $\&$ M University at Qatar, PO Box 23874, Doha, Qatar}
\affiliation{School of Computational Sciences, Korea Institute for Advanced Study, Seoul 130-012, Korea}
\author{Su-Yong Lee}
\affiliation{Department of Physics, Texas A $\&$ M University at Qatar, PO Box 23874, Doha, Qatar}
\author{Se-Wan Ji}
\affiliation{School of Computational Sciences, Korea Institute for Advanced Study, Seoul 130-012, Korea}
\author{M. S. Kim}
\affiliation{QOLS, Blackett Laboratory, Imperial College London, London SW7 2BW, United Kingdom}
\date{\today}
\begin{abstract}
We present a formalism to derive entanglement criteria beyond the Gaussian regime that can be readily tested by only homodyne detection.
The measured observable is the Einstein-Podolsky-Rosen (EPR) correlation. Its arbitrary functional form enables us to detect non-Gaussian entanglement even when an entanglement test based on second-order moments fails. We illustrate the power of our experimentally friendly criteria for a broad class of non-Gaussian states under realistic conditions. We also show rigorously that quantum teleportation for continuous variables employs a specific functional form of EPR correlation.
\end{abstract}
\maketitle

Quantum entanglement plays a key role in making quantum-mechanical predictions distinct from their classical counterparts. Its characterization and detection have been a topic of crucial importance that has attracted a great deal of effort, both theoretically and experimentally.
Despite remarkable progress, there still exists a pressing demand for further developments particularly for continuous variables (CVs).
In the CV regime, Gaussian entangled states have been the subject of study, but now considerable attention has been directed to non-Gaussian states.
This may be attributed to the fact that non-Gaussian entangled states not only provide practical advantages over their Gaussian counterparts, e.g. in quantum teleportation \cite{Opatrny} and dense coding \cite{Kitagawa2}, but also turn out to be essential, e.g. for universal quantum computation \cite{Lloyd} and nonlocality test \cite{Nha}.

The quantum state of a CV system is fully described in phase space by representing two quadrature variables $\hat{X}$ (position) and $\hat{P}$ (momentum).
To address the correlation of two CV systems, the EPR operators $\hat{u}'=|g|\hat{X}_A-{1\over g}\hat{X}_B$ and $\hat{v}'=|g|\hat{P}_A+{1\over g}\hat{P}_B$ can be defined, as Einstein, Podolsky, and Rosen (EPR) originally envisioned \cite{EPR}. Here $g$ is an arbitrary real number and $A$ and $B$ label each subsystem.
When the sum of the variances, $E_1'=\langle\Delta^2\hat{u}'\rangle+\langle\Delta^2\hat{v}'\rangle$, is smaller than $\frac{1}{2}\left(g^2+\frac{1}{g^2}\right)$, the state of the total system is inseparable~\cite{Duan} and such states are said to be EPR-correlated. In fact, it has been shown that the EPR correlation, after local unitary operations, is a necessary and sufficient condition for two-mode Gaussian entanglement~\cite{Duan} (In this Letter, we study a simple case, where we use $\{\hat{u},\hat{v}\}$ instead of $\{\hat{u}',\hat{v}'\}$ assuming $g=1$). Using the uncertainty principle as a requirement of a physical system in conjunction with the partial transposition \cite{Peres}, Simon found a necessary and sufficient condition for Gaussian entangled states (Simon criterion)~\cite{Simon}; this criterion is also concerned with the second moments of the quadrature variables~\cite{Giedke}. These criteria are theoretically easy to calculate and readily tested experimentally since the quadrature variables can be measured using highly-efficient homodyne detectors.

Beyond the Gaussian regime, some entanglement criteria have also been proposed \cite{Hillery, Agarwal, nha1,Shchukin,nha2,CMC,Walborn1,Walborn2}. Shchukin and Vogel particularly derived a hierarchy of entanglement conditions using higher-order moments \cite{Shchukin,nha2}.
For finite-dimensional systems, a remarkable criterion based on covariance matrices that unifies other criteria as its corollaries has been developed \cite{CMC}.  However, these tools are not necessarily practical for infinite-dimensional CV states because they may be nontrivial to assess theoretically and implement experimentally \cite{Shchukin1}.
Some criteria are at least theoretically easier to assess but they do not seem useful for a large class of non-Gaussian states.
Therefore, new entanglement criteria beyond the Gaussian regime must be developed, desirably in an experimentally friendly form, which is the objective of the current work.

Here we present a formalism employing the usual EPR operator $\hat{O}_{\rm EPR}\equiv\hat{u}^2+\hat{v}^2$ in an arbitrary functional form, which is readily testable with highly efficient homodyners and is theoretically convenient to calculate. These functional EPR criteria turn out to detect CV entanglement well beyond the Simon criterion, thereby proving genuine non-Gaussian entanglement.
Our criterion can be efficiently tested under realistic conditions including environmental interactions and detector inefficiency.
We illustrate the power of our formalism by considering a broad class of non-Gaussian states and show that it can be a crucial tool in addressing CV problems of fundamental importance, e.g. the robustness of Gaussian versus non-Gaussian entangled states against decoherence.

In order to present the new entanglement criterion, we briefly revisit the so-called Braunstein-Kimble (BK) protocol which is the standard for CV teleportation~\cite{Braunstein}. Unlike the original description, we study the protocol in the Heisenberg picture in order to gain a crucial insight. Let $\hat{a}_{\rm in}$ be an unknown input which is to be teleported, and $\hat{a}_A$ and $\hat{a}_B$ the entangled quantum channel sent to Alice and Bob, respectively.
Each field can be decomposed into real and imaginary parts (two quadrature amplitudes), $\hat{a}_i=\hat{X}_i+i\hat{P}_i$ ($i={\rm in, A, B})$.
In the BK protocol, Alice first superposes her mode with the input at a 50:50 beam splitter to obtain two output modes, $\hat{a}_{\pm}=(\hat{a}_{\rm in}\pm\hat{a}_A)/\sqrt{2}$, and measures the $\hat{X}$ ($\hat{P}$) quadrature of mode $\hat{a}_-$ ($\hat{a}_+$).
The measurement outcomes $\{X_{-}, P_{+}\}$ are transmitted to Bob who performs the displacement $\hat{D}(\delta)\equiv e^{\delta \hat{a}^\dag-\delta^*\hat{a}}$ with $\delta=\sqrt{2}(X_{-}+iP_{+})$ \cite{note}.  Although $X_{-}$ and $P_{+}$ are classical numbers, if formally represented by quantum operators \cite{Noh},
Bob's output becomes
\begin{eqnarray}
\hat{a}_{\rm out}=\hat{a}_{B}+\sqrt{2}\left({\hat{X}_{-}+i\hat{P}_{+}}\right)=\hat{a}_{\rm in}-\hat{a}^\dag_A+\hat{a}_B.
\label{eqn:HT1}
\end{eqnarray}
We rigorously justify Eq.~(\ref{eqn:HT1}) by showing that it leads to exactly the same mapping in the Schr{\" o}dinger picture \cite{Marian} (Supplemental Material).

{\it Teleportation fidelity}---As a figure of merit to assess the performance of the BK protocol, the output fidelity $F$ averaged over all input coherent states $|\beta\rangle=\hat{D}(\beta)|0\rangle$ ($\beta$: amplitude) can be used. Noting that the mean amplitudes of input and output fields always agree in the BK protocol \cite{note}, we realize that it suffices to look into the fidelity for the vacuum state input $|0\rangle$.
Let ${\hat U}_t$ be the global unitary operator that gives the output state $\rho_{\rm out}={\rm Tr}_{\rm A,B}\{{\hat U}_t|0\rangle\langle0|_{\rm in}\otimes\rho_{\rm AB}{\hat U}_t^\dag\}$ after teleportation via a quantum channel $\rho_{\rm AB}$ (Supplemental Material). The fidelity is given by $F=\langle0|\rho_{\rm out}|0\rangle={\rm Tr}\{\rho_{\rm out}|0\rangle\langle0|\}={\rm Tr}\{\rho_{\rm out}:e^{-\hat{a}_{\rm out}^\dag \hat{a}_{\rm out}}:\}$ with the normal-ordered expansion of the vacuum $|0\rangle\langle0|=:e^{-\hat{a}^\dag \hat{a}}:$ \cite{Barnett}. Using Eq.~(\ref{eqn:HT1}) and the vanishing contribution of the input vacuum state under normal-ordering, the output fidelity is reduced to an average over the quantum channel $\rho_{\rm AB}$,
\begin{eqnarray}
F=\sum_{m=0}^\infty\frac{(-1)^m}{m!}\langle \hat{a}^{\dag m}_{\rm out}\hat{a}^m_{\rm out}\rangle
=\left\langle e^{-\Delta^2\hat{u}-\Delta^2\hat{v}}\right\rangle_{\rho_{\rm AB}}
\label{eqn:fidelity}
\end{eqnarray}
using the identity $(\hat{a}_B^\dag-\hat{a}_A)(\hat{a}_B-\hat{a}_A^\dag)=\Delta^2\hat{u}+\Delta^2\hat{v}$ \cite{note}.
We see now that all orders of EPR correlations $E_m\equiv \langle(\Delta^2\hat{u}+\Delta^2\hat{v})^m\rangle$ contribute to the fidelity  nontrivially, and this can be used to address various issues on quantum teleportation using non-Gaussian channels.  As an example, due to an inequality $e^{-a^2}\ge1-a^2$, we deduce $F\ge1-E_1$. Therefore, the condition $E_1<\frac{1}{2}$ alone, using Gaussian or non-Gaussian channels, guarantees the fidelity over the classical limit $\frac{1}{2}$ \cite{Hammerer}.

For a Gaussian channel, higher-order correlations are constrained by the lowest one $E_1$, and $E_1<1$ becomes a necessary condition for faithful teleportation. Surprisingly, using the formulation above, we find that for a non-Gaussian channel, it is possible to have near-unit fidelity even without the usual EPR correlation, $E_1\ge1$. Therefore, the EPR correlation is not a necessary condition for a successful teleportation in the non-Gaussian regime.
Remarkably, beyond the teleportation context, this suggests {\it a possibility of detecting genuine non-Gaussian entanglement}, for which the Simon criterion fails. Note that (i) the fidelity above 1/2 proves quantum entanglement \cite{Hammerer} and (ii) the fidelity can be predicted also by directly measuring $E_{1j}$ of the quantum channel over $N$ runs, as  $F=\left\langle e^{-\Delta^2\hat{u}-\Delta^2\hat{v}}\right\rangle=\frac{1}{N}\sum_{j=1}^{N} e^{-E_{1j}}$. (See also {\it Implementation} below and Fig.~1~(a).)

{\it Entanglement criteria}---We present a formalism to construct an arbitrary functional form of EPR correlation as an entanglement criterion.
Given a general function ${\cal F}$, we consider an ensemble average of the Hermitian operator ${\cal F}( \hat{O}_{\rm EPR})$,
with $\hat{O}_{\rm EPR}\equiv\hat{u}^2+\hat{v}^2$.
As detailed in Supplemental Material, for separable states, the value of $\langle {\cal F}( \hat{O}_{\rm EPR})\rangle$ is bounded,
\begin{eqnarray}
{\cal F}_{\rm min}\le \langle{\cal F}( \hat{O}_{\rm EPR})\rangle  \le{\cal F}_{\rm max}.
\label{eqn:criterion}
\end{eqnarray}
Here the upper (lower) bound for separability is given by ${\cal F}_{\rm max (min)}\equiv \max(\min)_n\{O_n\}$ over $n=0,1,\dots$,
where
\begin{eqnarray}
O_n=4 \int_0^\infty dx dy{\cal F}(2x^2) e^{-\frac{1}{2}y^2} L_n(y^2)J_0(2xy) xy
\label{eqn:bound3}
\end{eqnarray}
($L_n$: Laguerre polynomial, $J_0$: Bessel function).
Eq.~(\ref{eqn:bound3}) is readily integrated for a regular function ${\cal F}$.
If the inequality~(\ref{eqn:criterion}) is violated for any function ${\cal F}$, the state is entangled.

Our method, when applied to the lowest-order EPR correlation $E_1'$, recovers Duan {\it et al.}'s criterion.
For higher-order criteria, using ${\cal F}(z)=z^m$,
we obtain separable conditions $E_m\equiv \langle(\Delta^2\hat{u}+\Delta^2\hat{v})^m\rangle\ge m!$.
Below we adopt a general ${\cal F}$ in the Taylor-series form and demonstrate its usefulness in detecting CV entangled states broadly.

{\it Implementation}---Separable states must satisfy the inequality~(\ref{eqn:criterion}) also under any local unitary transformations. This reasoning leads to a comprehensive set of entanglement criteria based on the EPR operator using generalized orthogonal quadratures. That is, the quadrature amplitudes under rotation, $\hat{X}_i'=\cos\phi_i\hat{X}_i-\sin\phi_i\hat{P}_i$ and $\hat{P}_i'=\sin\phi_i\hat{X}_i+\cos\phi_i\hat{P}_i$ $(i=A,B)$ can be employed and the violation of~(\ref{eqn:criterion}) for any single pair $\{\phi_A,\phi_B\}$ is sufficient to verify entanglement.
Given an entangled state, the violation usually occurs over a range of values $\phi_A$ and $\phi_B$ for a fixed ${\cal F}$, making an experimental test robust.

In experiments, the generalized $\hat{O}_{\rm EPR}$ can be measured by first phase-shifting each mode with $\phi_i$ $(i=A,B)$ and injecting them to a 50:50 beam-splitter (Fig. 1 (a)). One then measures the fixed quadrature $\hat{X}_1$ and $\hat{P}_2$ at each output, respectively, using homodyne detectors. ($\hat{X}_A'-\hat{X}_B'=\sqrt{2}\hat{X}_1$ and $\hat{P}_A'+\hat{P}_B'=\sqrt{2}\hat{P}_2$). We emphasize that neither a full state tomography nor the reconstruction of a probability distribution is needed for this test of entanglement: The EPR value $O_{{\rm EPR},j}$ obtained at each run ($j=1,\dots,N$) is plugged into a test function ${\cal F}$ to construct the average over $N$ runs as $\langle {\cal F}( \hat{O}_{\rm EPR})\rangle=\frac{1}{N}\sum_{j=1}^N {\cal F}(O_{{\rm EPR},j})$. It is then compared with the bounds in Eq. ~(\ref{eqn:bound3}) to verify entanglement.
As one obtains a finite number $N$ of data, it is of practical importance to see whether such a finite number of data suffices to beat the separability bounds properly. This issue can be addressed by looking into the statistical error $\delta_e$ of the mean, $\delta_e\sim\Delta_{\cal F}/\sqrt{N}$, where
\begin{eqnarray}
\Delta_{\cal F}^2\equiv\langle {\cal F}^2( \hat{O}_{\rm EPR})\rangle-\langle {\cal F}( \hat{O}_{\rm EPR})\rangle^2.
\label{eqn:errorbar}
\end{eqnarray}
We include this analysis below to demonstrate the power of our criteria under realistic conditions.

\begin{figure}
\centerline{\scalebox{0.35}{\includegraphics[angle=270]{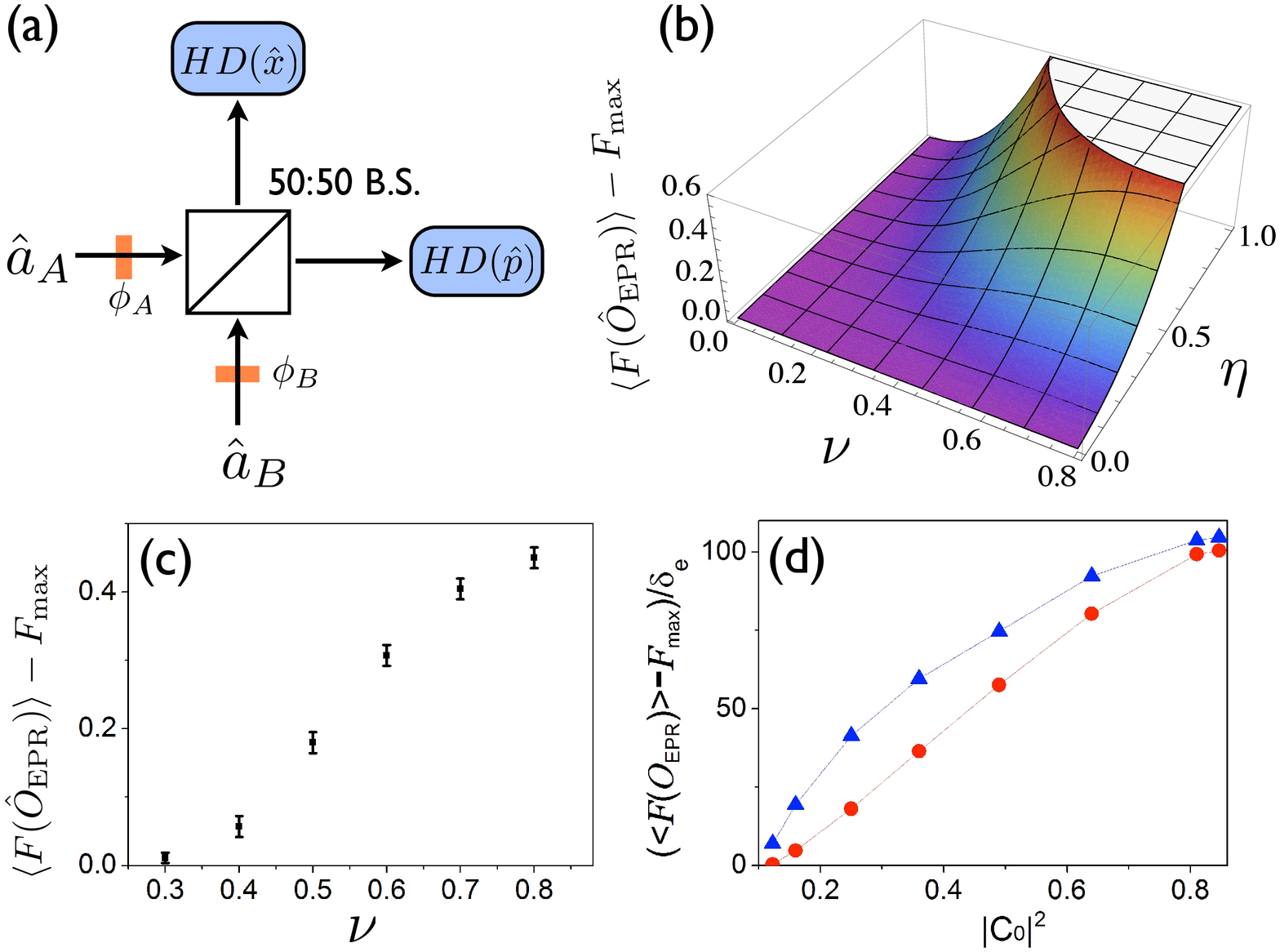}}}
\vspace{-0.2in}
\caption{(a) Experimental scheme to detect generic CV entanglement by only two homodyne detectors (HDs). $\phi_{A,B}$: phase shift. (b) $\langle {\cal F}( \hat{O}_{\rm EPR})\rangle-{\cal F}_{\rm max}$ as a function of $\nu$ (coherent amplitude) and $\eta$ (overall efficiency including transmission loss and detector efficiency) for a dephased cat state ($p$=0.3), with ${\cal F}(z)=e^{-Cz} (1+Dz)$.
At each point $\{\nu,\eta\}$, the positive value, which verifies entanglement, was optimized over the parameters $C\in(0,10]$ and $D\in[-80,80]$.
(c) $\langle {\cal F}( \hat{O}_{\rm EPR})\rangle-{\cal F}_{\rm max}$ as a function of $\nu$  ($p=0.5$), with the error bar $\delta_e=\Delta_{\cal F}/\sqrt{N}$ in Eq.~(\ref{eqn:errorbar}). (d) $(\langle {\cal F}( \hat{O}_{\rm EPR})\rangle-{\cal F}_{\rm max})/\delta_e$ for a state $c_0|00\rangle+c_1|11\rangle$ with ${\cal F}=e^{-Cz}$ (circle) and ${\cal F}=e^{-Cz} (1+Dz)$ (triangle).  In (c) and (d), $\eta=0.7$ and $n_{\rm th}=0.07$ (electronic noise) with $N=10^5$.}
\label{fig:fig1}
\end{figure}

{\it Trial functions}---For an entanglement test, we employ a function ${\cal F}(z)=e^{-Cz}\left(1+\sum_{m=1}^M D_m z^m\right)$, an exponentially decreasing function multiplied by a finite polynomial. This form can be regarded as the most general in view of the Taylor-series expansion of a regular function ($M\rightarrow\infty$). Our criterion becomes more powerful by adopting higher-order polynomials; Given an entangled state, our task is to find the set of parameters $\{C,D_1,\cdots,D_M\}$ for which the inequality~(\ref{eqn:criterion}) is violated. If it occurs at a level $M$, so does it at higher levels $M'>M$.
The same logic also applies to the degree of violation measured by, e.g., $(\langle {\cal F}( \hat{O}_{\rm EPR})\rangle-{\cal F}_{\rm max})/\delta_e$ optimized over the parameters $\{C,D_1,\cdots,D_M\}$.
That is, the larger the $M$ value, the higher the confidence level of the test for a given number $N$ of homodyne measurements [Fig. 1 (d)].
In this Letter we illustrate our method by confining the polynomial to the first order, ${\cal F}(z)=e^{-Cz} (1+Dz)$.
Remarkably, a broad class of non-Gaussian entangled states of current theoretical and experimental relevance can be detected with it.

Using ${\cal F}(z)=e^{-Cz} (1+Dz)$, Eq. ~(\ref{eqn:bound3}) gives
\begin{eqnarray}
O_n=\frac{(1-C)^{n-1}}{(1+C)^{n+2}} \left[1-C^2+D(1-C+2n)\right],
\label{eqn:bound30}
\end{eqnarray}
whose maximum is relevant for entanglement detection.

{\bf Case (i)}  $D=0$: Max $\{O_n\}$ is $\frac{1}{1+C}$ when $n=0$. Therefore, if there exists any $C>0$ such that $\langle e^{-C\hat{O}_{\rm EPR}}\rangle>\frac{1}{1+C}$, the state is entangled. For example, with $C=1$, the separability bound becomes 1/2, which is known as the classical limit for teleportation \cite{Hammerer}.

{\bf Case (ii)} $D\ne0$: In this case, the number $n$ yielding maximum $O_n$ in Eq.~(\ref{eqn:bound30}) depends on $C$ and $D$. It should thus be obtained case by case, which is still straightforward to calculate. For a given state, the violation of ~(\ref{eqn:criterion}) usually occurs in a broad range of $C$ and $D$, which makes our scheme robust against noise. Below we consider possible experimental imperfections like channel loss or thermal noise during signal transmission and detector inefficiency or electronic noise in homodyne detection. These errors can be modeled by
\begin{eqnarray}
\hat {a}_{i,d}=\sqrt{\eta}\hat {a}_{i,o}+\sqrt{1-\eta}\hat {v}_i \hspace{0.5cm} (i=A,B),
\label{eqn:emodel}
\end{eqnarray}
where $\hat{a}_{i,d}$ is the mode actually detected, $\hat{a}_{i,o}$ the original mode, and $\hat {v}_i$ the vacuum or thermal mode characterizing noise. The results below are all obtained analytically.
Our method is also compared with other CV criteria directly testable by homodyne detection such as the Simon criterion \cite{Simon} and the entropic criterion \cite{Walborn1,Walborn2}.

{\it Examples}---Let us first consider a class of dephased cat-states \cite{Grangier}, $\rho_{\rm cat}\sim|\nu,\nu\rangle\langle\nu,\nu|+|-\nu,-\nu\rangle\langle-\nu,-\nu|-p\left(|\nu,\nu\rangle\langle-\nu,-\nu|+|-\nu,-\nu\rangle\langle\nu,\nu|\right)$, where $|\nu\rangle$ is a coherent state of amplitude $\nu$ and $1-p>0$ represents the degree of dephasing. They are entangled for all $\nu>0$ and $p>0$, which is not detected by the Simon criterion. The entropic criterion detects entanglement only for a large $\nu$ and $p$, despite the improvement with the generalized Renyi entropy \cite{Walborn1,Walborn2}. In contrast, our criterion detects entanglement for any $\nu>0$ and $p>0$, with $\phi_A=\phi_B=0$ in Fig.~1 (a).  Considering further the overall efficiency $\eta$ characterizing channel loss and detector efficiency, we plot the value of $\langle {\cal F}( \hat{O}_{\rm EPR})\rangle-{\cal F}_{\rm max}$ in Fig.~1 (b), where each point is optimized over the parameters $C\in(0,10]$ and $D\in[-80,80]$ for a highly dephased state ($p=0.3$). The positive values in {\it all} ranges imply the detection of non-Gaussian entanglement with the degree of violation increased with $\nu$ and $\eta$. Moreover, let us address a realistic situation where a total number $N$ of data is obtained from homodyne detection with detector efficiency $\eta$ and electronic noise, which can be described by mixing signal with a thermal state ($n_{\rm th}$: thermal excitation number) \cite{Grangier2}.
In Fig.~1 (c), the case of dephased cat ($p=0.5$) is plotted as a function of $\nu$ for $\eta=0.7$ and $n_{\rm th}=0.07$ (electronic noise), with the error bar. The positive values above zero clearly verify entanglement.

Secondly, let us consider a state $|\Psi\rangle_{\rm B}=c_0|00\rangle+c_1|11\rangle$. We have checked that the Simon criterion and our method detect entanglement for any $c_0$ and $c_1$ values, whereas the entropic method \cite{Walborn2} does so for only $|c_0|>0.503$.
In Fig. 1 (d), we plot $(\langle {\cal F}( \hat{O}_{\rm EPR})\rangle-{\cal F}_{\rm max})/\delta_e$ for the state $|\Psi\rangle_{\rm B}$, which is improved by changing the test function from ${\cal F}=e^{-Cz}$ to ${\cal F}=e^{-Cz} (1+Dz)$.
Furthermore, suppose that the state undergoes decoherence through a thermal reservoir. The reservoir interaction is described by Eq.~(\ref{eqn:emodel}).
As $\eta^{-1}$ gets larger the system becomes more decohered. As shown in Fig.~2 (a), entanglement detection is significantly improved over the Simon method (square), particularly by modifying the test function from ${\cal F}(z)=e^{-Cz}$ (circle) to ${\cal F}(z)=e^{-Cz} (1+Dz)$ (triangle). Similar results are also obtained for higher-dimensional entangled states.

Finally, the classes of entangled states generated by practically available non-Gaussian operations, i.e., photon-subtraction \cite{Grangier} and addition \cite{Bellini} are of current interest~\cite{Kim}. When these operations are applied to a two-mode squeezed state $|{\rm TMSS}\rangle=e^{s(\hat{a}_A^\dag\hat{a}_B^\dag-\hat{a}_A\hat{a}_B)}|00\rangle$, the Simon criterion fails to verify the entanglement of the photon-added state $\hat{a}_A^\dag\hat{a}_B^\dag|{\rm TMSS}\rangle$ for $s<0.378$, whereas our method detects entanglement for the whole range of $s>0$. Furthermore, we address a question of fundamental importance raised recently by Allegra {\it et al.}, i.e., whether Gaussian entanglement is more robust than non-Gaussian entanglement against decoherence due to the thermal reservoir \cite{Paris}. We show that non-Gaussian entanglement can be more robust than Gaussian entanglement in the parameter regime they studied \cite{nha3}. The conclusion of \cite{Paris} was drawn by using several previously known criteria for verifying non-Gaussian entanglement, which however did not outperform the Simon criterion. In contrast, our method analytically shows that the photon-subtracted states (triangle: our criterion, square: Simon criterion) survive longer under thermal noise than their Gaussian counterparts of the same initial energy (circle) [Fig. 2 (b)]. Similar results are also obtained for the photon added states.

\begin{figure}
\centerline{\scalebox{0.3}{\includegraphics[angle=0]{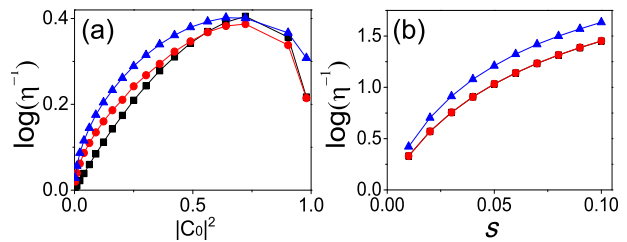}}}
\vspace{-0.8in}
\caption{ Entanglement is detected till the decoherence time $\log\eta^{-1}$ for the initial state (a) $c_0|00\rangle+c_1|11\rangle$ and (b) $\hat{a}_A\hat{a}_B|{\rm TMSS}\rangle$ interacting with a thermal reservoir of (a) $n_{\rm th}=0.5$ and (b) 0.05, respectively. Triangles: our method with ${\cal F}(z)=e^{-Cz} (1+Dz)$, squares: Simon criterion, and circles: (a) our method with ${\cal F}(z)=e^{-Cz}$ (b) Simon criterion for $|{\rm TMSS}\rangle$ having the same initial energy as $\hat{a}_A\hat{a}_B|{\rm TMSS}\rangle$.}
\label{fig:fig2}
\end{figure}



{\it Remarks}---We have developed a method to derive entanglement criteria beyond the Gaussian limit that can be efficiently tested by homodyne detection. We illustrated our method by adopting two trial functions which broadly detect non-Gaussian entanglements of fundamental and practical interests. We also showed that our method is robust against experimental imperfections. The trial functions can be made more powerful by incorporating higher-order polynomials, which do not however incur any extra experimental efforts. On another side, the question on how much useful (robust) non-Gaussian entangled states are under practical conditions is of importance for quantum information processing. Our study for quantum teleportation, clarified here as employing the EPR correlation in a specific form, shows that non-Gaussian channels can be more robust against decoherence than Gaussian ones.

H. N. acknowledges the support by the NPRP 4-520-1-083 from Qatar National Research Fund and M. S. K. the financial support by the UK EPSRC.


\section{Supplemental Material}
{\bf Part I}: We first show that Eq.~(1) is a legitimate transformation describing the CV quantum teleportation that leads to exactly the same mapping in the Schr{\" o}dinger picture [22]. In Ref. [21], Eq.~(1) was used for the study of teleporting temporal correlation functions by only noting that the operator $\hat{\delta}\equiv\sqrt{2}\left(\hat{X}_{-}+i\hat{P}_{+}\right)=\hat{a}_{\rm in}-\hat{a}^\dag_A$ acts like a classical number due to $[\hat{\delta},\hat{\delta}^\dag]=0$. In contrast, we rigorously justify Eq.~(1) as follows.

Note that $\hat{a}_{\rm out}=\sqrt{2}\frac{\hat{a}_{\rm in}+\hat{a}_B}{\sqrt{2}}-\hat{a}^\dag_A$ can be understood as superposing two modes $\hat{a}_{\rm in}$ and $\hat{a}_B$ at a 50:50 beam splitter followed by the nondegenerate parametric amplification (gain $G=2$) with the idler mode $\hat{a}_A$.
This suggests the conjugate evolutions for $\hat{a}_A$ and $ \hat{a}_B$,
\begin{eqnarray}
\hat{a}_{A}'=\sqrt{2}\hat{a}_A-\frac{\hat{a}^\dag_{\rm in}+\hat{a}^\dag_B}{\sqrt{2}},\hspace{1.2cm}
\hat{a}_B'=\frac{\hat{a}_{\rm in}-\hat{a}_B}{\sqrt{2}}.\setcounter{equation}{8}
\label{eqn:HT2}
\end{eqnarray}
With the Heisenberg transformations in Eqs.~(1) and (8) taken as $\hat{a}'_i={\hat U}_t^\dag \hat{a}_i{\hat U}_t$ ($i={\rm in, A, B})$, where ${\hat U}_t$ is a global unitary operator corresponding to CV teleportation,
one can obtain the characteristic function of the output, $C_{\rm out}(\lambda)={\rm tr} \{ \hat{D}(\lambda)\rho_{\rm out}\}$,
where $\rho_{\rm out}={\rm tr}_{A,B} \{{\hat U}_t\rho_{\rm in}\otimes \rho_{AB}{\hat U}_t^\dag\}$, as
\begin{eqnarray}
C_{\rm out}(\lambda)=C_{\rm in}(\lambda)C_{\rm AB}(\lambda^*,\lambda),\setcounter{equation}{9}
\label{eqn:cha}
\end{eqnarray}
which is in agreement with [22].

{\bf Part II}: We next prove that there exist quantum channels that do not have the usual EPR correlation, i.e. $E_1\ge1$, but can teleport any input state with fidelity arbitrarily close to unity.
For simplicity, let us consider a class of entangled channels produced by injecting identical fields (with only a phase shift to the second) of modes $\hat{a}_1$ and $\hat{a}_2$ to a 50:50 beam splitter. The output modes
$\hat{a}_A=(\hat{a}_1+\hat{a}_2)/\sqrt{2}$ and $\hat{a}_B=(-\hat{a}_1+\hat{a}_2)/\sqrt{2}$ give the fidelity in Eq.~(2) as $F=\langle e^{-2\hat{X}_1^2}\rangle\langle e^{-2\hat{P}_2^2}\rangle$.
For high fidelity, we only need to maximize $f_1\equiv\langle e^{-2\hat{X}_1^2}\rangle$ for a single mode; $f_2\equiv\langle e^{-2\hat{P}_2^2}\rangle$ can be similarly maximized by applying $\frac{\pi}{2}$ phase-shift ($\hat{X}\rightarrow\hat{P}$). To ensure the no-EPR-correlation, $E_1\ge1$, for this channel, we impose the condition $\langle \hat{X}_1^2\rangle=\frac{1}{4}$ (no squeezing). Our task is now to make $\langle e^{-2\hat{X}_1^2}\rangle$ close to unity with $\langle \hat{X}_1^2\rangle=\frac{1}{4}$.
Obviously, the quadrature distribution $p(X)$ must be sharply centered around $X=0$ to obtain a large $\langle e^{-2\hat{X}^2}\rangle$,
but presumably with a rather long tail to simultaneously ensure a finite variance, $\langle \hat{X}^2\rangle=\frac{1}{4}$.
This leads us to envision a probability distribution that decreases with $t\equiv 2X^2$ more rapidly than the Gaussian function.
We can construct a number of such distributions analytically, e.g.  $p_m(t)\sim e^{-Ct^{\frac{1}{m}}}$ ($m>1$).
After normalization, we obtain a class of probability distributions
\begin{eqnarray}
p_m(X)=\frac{2(2m)!}{(m!)^2}|X| e^{-\left(\frac{2(2m)!}{m!}\right)^{\frac{1}{m}}|X|^{\frac{2}{m}}},\setcounter{equation}{10}
\label{eqn:nonGaussian}
\end{eqnarray}
and find that $f_1=\int dXp_m(X)e^{-2X^2}$ approaches unity as $m$ increases.

One can construct a corresponding non-Gaussian entangled state, $\rho_{AB}=U_{\rm BS}\rho_1\otimes\rho_2 U_{\rm BS}^\dag$ ($U_{\rm BS}$: beam-splitter)
with two single-mode states $\rho_1=\int dX p_m(X) |X\rangle\langle X|$ and $\rho_2=\int dP p_m(P) |P\rangle\langle P|$,
where $|X\rangle$ and $|P\rangle$ are $\hat{X}$- and $\hat{P}$-quadrature eigenstates with the probability distribution $p_m$ in Eq.~(10).
The uncertainty principle yields $\Delta^2 P \ge\frac{1}{16\Delta^2 X}=\frac{1}{4}$ for the state $\rho_1$. No squeezing is then assured for arbitrary quadratures $\hat{X}\cos\theta+\hat{P}\sin\theta$ with a little algebra.
Furthermore, it can be shown that the state $\rho_{AB}$ asymptotically gives the unit fidelity for all input states; Its characteristic function $C_{\rm AB}(\lambda^*,\lambda)=C_p(\lambda_x)C_p(\lambda_y)$, where $C_p(k)\equiv\int dXp_m(X) e^{-2\sqrt{2}ikX}$, becomes very flat over the entire range of $\lambda$ with $m$ increased, which ensures a high-fidelity via Eq.~(9).
As an example, in Fig.~3, we plot the Wigner function of the teleported state for the input Fock state $|1\rangle$.
The overall features are successfully teleported except for a relatively long tail in the output, whose effect on fidelity is however negligible.

\begin{figure}
\centerline{\scalebox{0.35}{\includegraphics[angle=270]{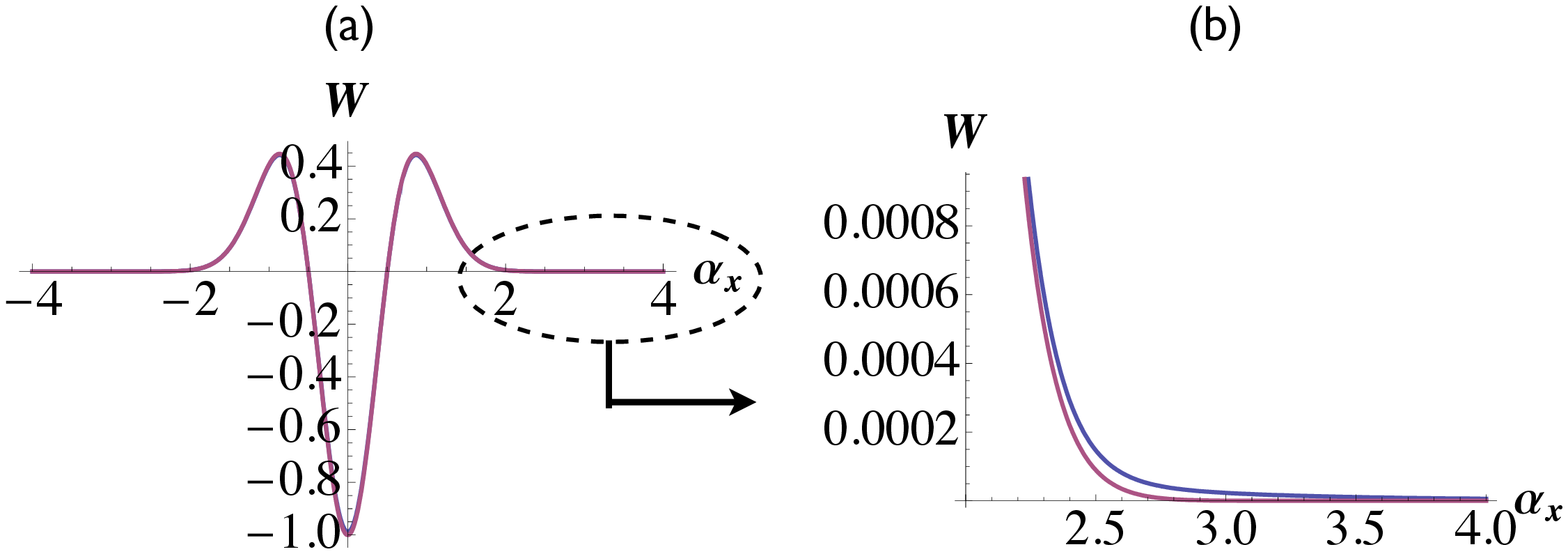}}}
\vspace{-1.45in}
\setcounter{figure}{2}
\caption{Wigner functions $W(\alpha_x,\alpha_y=0)$ for the input Fock state $|1\rangle$ and the teleported state (indiscernible with fidelity $F\approx0.992$) by the quantum channel in Eq.~(10) for $m=50$ (a). In (b) a rather long tail is shown in the output (upper blue curve), which does not significantly affect fidelity.}
\label{fig:fig3}
\end{figure}

{\bf Part III}: Finally, we show how the separability condition in (3) can be derived employing an arbitrary function ${\cal F}$.
It is well known that for a separable state, the density operator under partial transposition (PT) is still positive-semidefinite [7], thus satisfying all requirements imposed on a legitimate quantum state.
As demonstrated below, this leads to an upper/lower bound of the value $\langle {\cal F}( \hat{O}_{\rm EPR})\rangle$ that a separable state can take, where $\hat{O}_{\rm EPR}\equiv\hat{u}^2+\hat{v}^2=(\hat{X}_A-\hat{X}_B)^2+(\hat{P}_A+\hat{P}_B)^2$.
For each ${\cal F}$, one can find out such bounds as prescribed below, the violation of which is a sufficient condition for entanglement.

First, note that when $\hat{a}_A$ and $\hat{a}_B$ are input modes to a 50:50 beam splitter [Fig.~1 (a)], the operators $\hat{u}=\hat{X}_A-\hat{X}_B$ and $\hat{v}=\hat{P}_A+\hat{P}_B$ correspond to two commuting observables $\sqrt{2}\hat{X}_1$ and $\sqrt{2}\hat{P}_2$, respectively, at the output modes.
From the viewpoint of the output state, it is thus obvious that the Hermitian operator ${\cal F}( \hat{O}_{\rm EPR})$, for any function ${\cal F}$, can be regarded as a proper symmetric-ordered operator.
Therefore, one can generally express the quantum average $\langle {\cal F}( \hat{O}_{\rm EPR})\rangle$ by means of two-mode Wigner distribution $W_{AB}(\alpha,\beta)$
as
\begin{eqnarray}
&&\langle {\cal F}( \hat{O}_{\rm EPR})\rangle\nonumber\\&&=\int d^2\alpha d^2\beta W_{AB}(\alpha,\beta){\cal F}\left[(\alpha_x-\beta_x)^2+(\alpha_y+\beta_y)^2\right],\nonumber\\
\label{eqn:exp}
\end{eqnarray}
where the EPR operator $\hat{O}_{\rm EPR}=(\hat{X}_A-\hat{X}_B)^2+(\hat{P}_A+\hat{P}_B)^2$ is replaced by the corresponding classical amplitudes ($x$/$y$: real/imaginary parts).
Under PT, the Wigner function changes as $W_{AB}(\alpha_x,\alpha_y,\beta_x,\beta_y)\rightarrow W_{AB}(\alpha_x,\alpha_y,\beta_x,-\beta_y)$ [8] and Eq.~(\ref{eqn:exp}) becomes
\begin{eqnarray}
{\cal M}&&\equiv\langle {\cal F}( \hat{O}_{\rm EPR})\rangle_{\rm PT}\nonumber\\
&&=\int d^2\alpha d^2\beta W_{AB}(\alpha,\beta){\cal F}\left[(\alpha_x-\beta_x)^2+(\alpha_y-\beta_y)^2\right]. \nonumber\\
\label{eqn:bound1}
\end{eqnarray}
With the change of integration variables (redefining new modes) as $\alpha_1=\frac{\alpha-\beta}{\sqrt{2}}$ and $\alpha_2=\frac{\alpha+\beta}{\sqrt{2}}$, the above expression can be recast to
\begin{eqnarray}
{\cal M}=\int d^2\alpha_1 d^2\alpha_2  W_{12}(\alpha_1,\alpha_2){\cal F}\left(2|\alpha_1|^2\right)
\label{eqn:bound11}
\end{eqnarray}
where $W_{12}(\alpha_1,\alpha_2)=W_{AB}\left(\frac{\alpha_1+\alpha_2}{\sqrt{2}},\frac{\alpha_1-\alpha_2}{\sqrt{2}}\right)$ represents a certain two-mode Wigner function.
We now see that the extremal values of ${\cal M}$ for {\it all} quantum states are the same as those of a single-mode $W_1(\alpha_1)=\int d^2\alpha_2W_{12}(\alpha_1,\alpha_2)$ as $ {\cal F}$ now involves only one mode $\alpha_1$ in Eq.~(\ref{eqn:bound11}). From now on, we thus work with only a single-mode state $\rho$ having a Wigner function $W(\alpha)$.
Writing $W(\alpha)$ as the Fourier transform of the characteristic function $C(\lambda)={\rm Tr}\left[\rho {\hat D}(\lambda)\right]$, i.e.,
$W(\alpha)=\frac{1}{\pi^2}\int d^2\lambda C(\lambda)e^{\alpha\lambda^*-\alpha^*\lambda}$, we obtain
\begin{eqnarray}
{\cal M}=
\int d^2\alpha W(\alpha){\cal F}\left(2|\alpha|^2\right)
={\rm Tr}\left[\rho\hat{O}\right],
\label{eqn:bound2}
\end{eqnarray}
where
\begin{eqnarray}
\hat{O}\equiv\frac{1}{\pi^2}\int d^2\alpha d^2\lambda e^{\alpha\lambda^*-\alpha^*\lambda}{\hat D}(\lambda){\cal F}\left(2|\alpha|^2\right).
\end{eqnarray}
Now the problem is reduced to that of obtaining extremal expectation values of the operator $\hat{O}$ in Eq.~(\ref{eqn:bound2}).
Obviously, the maximum (minimum) value of ${\rm Tr}\left[\rho\hat{O}\right]$ is the maximum (minimum) eigenvalue of $\hat{O}$. One can show by a direct calculation that the operator $\hat{O}$ does not have off-diagonal elements in the number-state basis, i.e. $\langle m|\hat{O}|n\rangle=0$ for $m\ne n$. Thus, the eigenvalues of $\hat{O}$ are the diagonal elements $O_n\equiv\langle n|\hat{O}|n\rangle$ given by
\begin{eqnarray}
O_n=4 \int_0^\infty dx dy{\cal F}(2x^2) e^{-\frac{1}{2}y^2} L_n(y^2)J_0(2xy) xy
\label{eqn:bound33}
\end{eqnarray}
using $\langle n|{\hat D}(\lambda)|n\rangle=e^{-\frac{1}{2}|\lambda|^2}L_n(|\lambda|^2)$. ($L_n$: Laguerre polynomial, $J_0$: Bessel function)
The ensemble average ${\cal M}=\langle \hat{O}\rangle$ in Eq.~(\ref{eqn:bound2}) can thus take a value only in the range $[{\cal F}_{\rm min},{\cal F}_{\rm max}]$ where  ${\cal F}_{\rm min (max)}\equiv \min(\max)_n\{O_n\}$ over $n=0,1,\cdots$.
Therefore, for any function ${\cal F}$, a separable state must satisfy
\begin{eqnarray}
{\cal F}_{\rm min}\le \langle{\cal F}( \hat{O}_{\rm EPR})\rangle  \le{\cal F}_{\rm max}.
\label{eqn:criterion}
\end{eqnarray}


As an example, our method, applied to the lowest-order EPR correlation $E_1'\equiv \langle\Delta^2\hat{u}'\rangle+\langle\Delta^2\hat{v}'\rangle$, recovers Duan {\it et al.}'s criterion.
For this case, the argument of ${\cal F}$ in Eq.~(\ref{eqn:bound1}) becomes $(\alpha_i-\beta_i)^2\rightarrow (|g|\alpha_i-\frac{1}{g}\beta_i)^2$ $(i=x,y)$. Then, by redefining a single-mode $\alpha_1\equiv\frac{1}{\sqrt{g^2+1/g^2}}(|g|\alpha-\frac{1}{g}\beta)$, we similarly obtain the bounds, Eq.~(\ref{eqn:bound33}), with only the replacement ${\cal F}(2x^2)\rightarrow{\cal F}\left[(g^2+\frac{1}{g^2})x^2\right]$. Identifying ${\cal F}(z)=z$, one obtains the minimum $O_{n}=\frac{1}{2}(g^2+\frac{1}{g^2})$ with $n=0$. Therefore, we obtain the Duan {\it et al.}'s criterion as $E_1'\equiv \langle\Delta^2\hat{u}'\rangle+\langle\Delta^2\hat{v}'\rangle\ge\frac{1}{2}(g^2+\frac{1}{g^2})$.


\vspace{-0.5cm}
\begin{references}
\bibitem{Opatrny}T. Opatrny, G. Kurizki, and D.-G. Welsch, \pra {\bf 61}, 032302 (2000); P. T. Cochrane, T. C. Ralph, and G. J. Milburn,  {\it ibid.} {\bf 65}, 062306 (2002); S. Olivares, M. G. A. Paris, and R. Bonifacio, {\it ibid.} {\bf 67}, 032314 (2003); A. Kitagawa {\it et al.},  {\it ibid.} {\bf 73}, 042310 (2006); Y. Yang and F. L. Li, {\it ibid.} {\bf 80}, 022315 (2009); F. Dell'Anno {\it et al.}, {\it ibid.} {\bf 76}, 022301 (2007); F. Dell'Anno, S. De Siena, and F. Illuminati, {\it ibid.} {\bf 81}, 012333 (2010); S.-Y. Lee {\it et al.}, Phys. Rev. A 84, 012302 (2011).
\bibitem{Kitagawa2} A. Kitagawa {\it et al.}, \pra {\bf 72}, 022334 (2005).
\bibitem{Lloyd} S. Lloyd and S. L. Braunstein, \prl {\bf 82}, 1784 (1999); S. D. Bartlett and B. Sanders, {\it ibid.}, {\bf 89}, 207903 (2002).
\bibitem{Nha} H. Nha and H. J. Carmichael, \prl {\bf 93}, 020401 (2004);  R. Garcia-Patron {\it et al.}, {\it ibid.} {\bf 93}, 130409 (2004).
\bibitem{EPR} A. Einstein, B. Podolsky, and N. Rosen, Phys. Rev. {\bf 47}, 777 (1935).
\bibitem{Duan} L.-M. Duan {\it et al.}, \prl {\bf 84}, 2722 (2000).
\bibitem{Peres} A.~Peres, \prl {\bf 77}, 1413 (1996).
\bibitem{Simon} R.~Simon, \prl {\bf 84}, 2726 (2000).
\bibitem{Giedke} G. Giedke {\it et al.}, \prl {\bf 87}, 167904 (2001).
\bibitem{Hillery} M.~Hillery and M.~S.~Zubairy, \prl {\bf 96}, 050503 (2006).
\bibitem{Agarwal}G.~S.~Agarwal and A.~Biswas, New J.~Phys. {\bf 7}, 211 (2005).
\bibitem{nha1} H.~Nha and J.~Kim, \pra {\bf 74}, 012317 (2006): H. Nha, \pra {\bf 76}, 014305 (2007).
\bibitem{Shchukin} E.~Shchukin and W.~Vogel, \prl {\bf 95}, 230502 (2005).
\bibitem{nha2} See also an uncertainty-relation-based approach. H.~Nha and M.~S.~Zubairy, \prl {\bf 101}, 130402 (2008); Q. Sun, H. Nha, and M. S. Zubairy, \pra {\bf 80}, 020101 (2009).
\bibitem{CMC} O. G{\"u}hne {\it et al.}, \prl {\bf 99}, 130504 (2007); O. Gittsovich {\it et al.}, \pra {\bf 78}, 052319 (2008).
\bibitem{Walborn1}S. P. Walborn {\it et al.}, \prl {\bf 103}, 160505 (2009).
\bibitem{Walborn2} A. Saboia, F. Toscano, and S. P. Walborn, \pra {\bf 83}, 032307 (2011).
\bibitem{Shchukin1} E.~Shchukin and W.~Vogel, \prl {\bf 96}, 200403 (2006). This proposed a homodyne correlation measurement that can be potentially useful for higher-order moments, for which the device becomes increasingly large.
\bibitem{Braunstein} S. L. Braunstein and H. J. Kimble, \prl {\bf 80}, 869 (1998).
\bibitem{note} We here assume that both Alice's and Bob's modes have zero means, $\langle\hat{a}_{A}\rangle=\langle\hat{a}_{B}\rangle=0$, e.g. the case of a usual two-mode squeezed state serving as a quantum channel.
If not, Bob can adjust the required displacement to $\alpha'=\sqrt{2}(X_{-}+iP_{+})+\langle\hat{a}^\dag_{A}\rangle-\langle\hat{a}_{B}\rangle$
to satisfy the condition $\langle\hat{a}_{\rm out}\rangle=\langle\hat{a}_{\rm in}\rangle$.
\bibitem{Noh} C. Noh {\it et al.}, \prl {\bf 102}, 230501 (2009).
\bibitem{Marian}P. Marian and T. A. Marian, \pra {\bf 74}, 042306 (2006).
\bibitem{Barnett} S. M. Barnett and P. M. Radmore, {\it Methods in Theoretical Quantum Optics}, Oxford University Press (2003).
\bibitem{Hammerer} S. L. Braunstein, C. A. Fuchs, and H. J. Kimble, J. Mod. Opt. {\bf 47}, 267 (2000);
K. Hammerer {\it et al.}, \prl {\bf 94}, 150503 (2005).
\bibitem{Grangier} A. Ourjoumtsev {\it et al.}, Nature Physics {\bf 5}, 189 (2009).
\bibitem{Grangier2} J. Lodewyck {\it et al.}, \pra {\bf 76}, 042305 (2007).
\bibitem{Bellini} A. Zavatta, S. Viciani, and M. Bellini, Science {\bf 306}, 660 (2004);  G. S. Agarwal and K. Tara, \pra {\bf 43}, 492 (1991).
\bibitem{Kim} M. S. Kim, J. Phys. B {\bf 41}, 133001 (2008).
\bibitem{Paris} M. Allegra, P. Giorda, and M. G. A. Paris, \prl {\bf 105}, 100503 (2010).
\bibitem{nha3} See also K. K. Sabapathy, J. S. Ivan, and R. Simon, \prl {\bf 107}, 130501 (2011); J. Lee, M. S. Kim, and H. Nha, \prl {\bf 107}, 238901 (2011).
\end{references}
\end{document}